\documentclass[%
 reprint,longbibliography,preprintnumbers,
nofootinbib,
 amsmath,amssymb,
 aps,
prl,
]{revtex4-1}
\pdfoutput=1
\usepackage[utf8]{inputenc}
\usepackage{flushend}
\usepackage{dcolumn}
\usepackage{bm}
\usepackage{balance}

\usepackage[normalem]{ulem}

\usepackage[colorlinks = true,
            linkcolor = blue,
            urlcolor  = blue,
            citecolor =green,
            anchorcolor = blue]{hyperref}
\usepackage{verbatim}
\usepackage{color,ulem}
\usepackage[english]{babel}

\usepackage[utf8]{inputenc}
\input Starburst.fd
\newcommand*\initfamily{\usefont{U}{Starburst}{xl}{n}}\initfamily

\newcommand{\beq}{\begin{eqnarray}}
\newcommand{\eeq}{\end{eqnarray}}
\usepackage{amsmath}
\usepackage{tikz}
\usetikzlibrary{decorations.pathmorphing}
\usetikzlibrary{shapes.misc}
\tikzset{cross/.style={cross out, draw=black, minimum size=8*(#1-\pgflinewidth), inner sep=0pt, outer sep=0pt},
cross/.default={1pt}}
\usetikzlibrary{patterns,math}
\begin{document}

\title{Theory of heavy-quarks contribution to the quark-gluon plasma viscosity}

\author{\textbf{Alessio Zaccone$^{1,2}$}}%
 \email{alessio.zaccone@unimi.it}

 \vspace{1cm}
 
\affiliation{$^{1}$Department of Physics ``A. Pontremoli'', University of Milan, via Celoria 16,
20133 Milan, Italy.}

\affiliation{$^{2}$Institute of Theoretical Physics, University of G\"ottingen, Friedrich-Hund-Platz 1, 37077 G\"ottingen, Germany.}

\begin{abstract}
The shear viscosity of quark gluon plasma is customarily estimated in the literature using kinetic theory, which, however, is well known to break down for dense interacting systems. Here we propose an alternative theoretical approach based on recent advances in the physics of dense interacting liquid-like systems, which is valid for strongly-interacting and arbitrarily dense relativistic systems.
With this approach, the viscosity of strongly interacting dense heavy-quarks plasma is evaluated analytically, at the level of special relativity. For QGP well above the confinement temperature, the theory predicts that the viscosity increases with the cube of temperature, in agreement with evidence. 
\end{abstract}

\maketitle

Understanding shear viscosity at a microscopic level is vital to solving outstanding mysteries in physics, such as the so-called viscosity minimum \cite{Murillo} and 
the existence of nearly non-dissipative fluids described by the
Euler hydrodynamics such as quark-gluon plasmas \cite{quark,10.21468/SciPostPhys.10.5.118,shen2023} and the microscopic mechanism of superfluidity in helium-4 \cite{Trachenko_2023}. 

Relativistic Heavy Ion Collider
(RHIC) experiments at Brookhaven National Laboratory studying Au + Au collisions provided evidence of a new state of matter, the quark-gluon plasma (QGP), at extremely high temperature and density, where hadrons melt into a soup of their de-confined constituents, i.e. quarks and gluons \cite{review}.
Surprisingly, the experimental data on QGP were initially found to be well fitted by hydrodynamic models assuming Euler inviscid flow. Such nearly dissipationless hydrodynamics is only possible when the shear viscosity $\eta$ is very low. Finding a mechanistic explanation to this surprising result has been a grand challenge to theoretical physics for the last 30 years \cite{Shuryak}. 

In parallel, calculation of the dimensionless ratio of shear viscosity $\eta$ to entropy density $s$ by Kovtun, Son and Starinets (KSS) within anti-de Sitter/conformal field theory (AdS/CFT) gave a lower bound on the viscosity $\eta/s \geq \hbar/4\pi k_{B}$.
This has been believed to be a universal lower bound to be obeyed by all physical systems, although violations of this bound have been reported  \cite{violation,Gochan_2019,PhysRevB.92.125103}, including, possibly, in the QGP viscosity \cite{Romatschke}.
Recent Bayesian analysis of experimental data support values of the QGP viscosity that are comparable to the KSS bound \cite{Bernhard2019}.

Much effort has been devoted to computing the viscosity of QGP numerically and from first-principles theoretical approaches \cite{Nakamura}. 
Most recent theoretical models based on AdS/CFT \cite{Noronha_2022} confirm the existence of a nearly dissipationless perfect fluid regime at high densities. However, effective field theory approaches, such as those based on AdS/CFT, cannot provide a microscopic mechanistic understanding of how viscosity vanishes based on microscopic interactions.
Other approaches widely used in the literature are based on kinetic theory, i.e. the Boltzmann equation with the Chapman-Enskog expansion in the relaxation time approximation \cite{PhysRevC.101.045203,Kapusta}. 
As is well known in statistical physics \cite{Balescu}, kinetic theory based on the Chapman-Enskog framework can, at best, provide a semi-quantitative description of the viscosity of gases, and even with the most modern corrections the theory is not fully quantitative \cite{Chapman_2023}, let alone for liquids.
In practice, the kinetic theory (Chapman-Enskog) is applicable only when the dimensionless parameter $r_c^{3}\,n$ is small, where $r_c$ is the interaction range between the particles and $n$ is the number density \cite{Balescu}. Considering the typical mass of a quark (3.56 - 8.9 $\times 10^{-30}$ kg) and the typical interaction range (1 fm), the dimensionless parameter is in the range $r_c^{3}\,n
\sim 10^3-10^4$, which is certainly not small and thus makes the kinetic theory inapplicable. 
 
Alternatively, the Green-Kubo formalism based on time correlation functions of the stress-energy tensor  \cite{Kaiser,Borsanyi}, appears to be the best suited approach, although for calculations of real systems it involves the fitting of long-time tails of autocorrelation functions, which is an art in its own. Also, it does not give access to simple analytical formulae.  Among recent approaches that focused on the diffusivity of heavy quarks, it is worth mentioning the nonperturbative treatment of Ref. \cite{Liu2020}. Recent state-of-art calculations of QGP viscosity using the Green-Kubo formalism within lattice QCD can be found in Ref. \cite{altenkort2022viscosity}. Previous calculations \cite{Astrakhantsev2017} based on measuring correlations of the energy-momentum tensor within lattice gluodynamics found viscosity to entropy-density ratios close to the KSS bound.

Here, we present a new method of computing the viscosity of dense strongly-correlated matter based on combining the concept of nonaffine displacements in dense disordered systems \cite{Scossa,zaccone2023} with the Langevin transport model of heavy quarks \cite{Teaney_2005,Teaney_2006}.

\begin{figure}[h]
\centering
\includegraphics[width=0.9\linewidth]{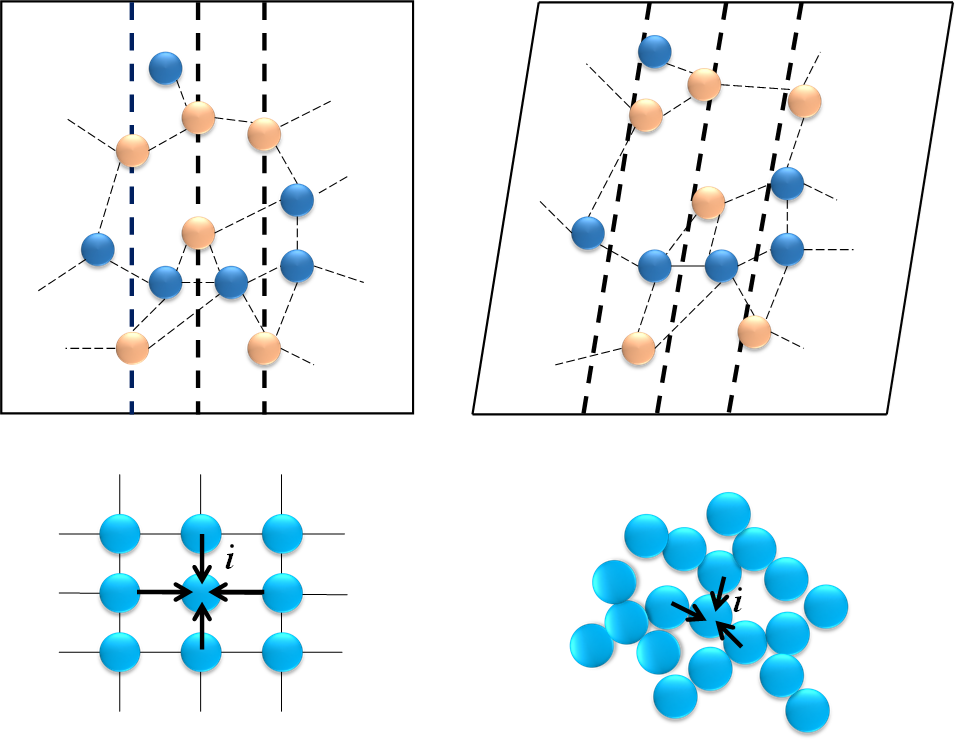}
\caption{The top panel shows particles in a strongly interacting dense fluid subjected to a shear deformation. Particles that, in the undeformed configuration (left) sit on the thick dashed vertical lines would land on the thick dashed lines of the deformed configuration (right) if the deformation were completely affine. Since the deformation of a disordered dense system is always nonaffine due to the lack of centrosymmetry around a tagged particle, these particles are displaced with respect to the thick dashed lines in the sheared configuration. The distance between these actual positions and the affine positions on the thick dashed lines represent the nonaffine displacements.
The lower panel illustrates the fact that in a centrosymmetric system such as a crystal lattice, all forces acting on a tagged particle in the affine position would cancel each other out by centrosymmetry (left panel) whereas in a dense disordered system such as a QGP the tagged particle in the affine position is not occupying a center of inversion symmetry, which means that the net force transmitted from its nearest neighbours is not zero. This net force has to be relaxed via an additional (nonaffine) displacement (see top panel).}
\label{fig1}
\end{figure}

We start from basic viscoelastic theory (which applies to both solids and liquids \cite{landau1959fluid}), by introducing the complex shear modulus $G^{*}$ \cite{Landau_elasticity,landau1959fluid,lakes_2009}.
The real and imaginary part, $G'$ and $G''$, of $G^{*}$ correspond to the dissipationless and to the dissipative part of the response, respectively, and they are also known as the storage (elastic) modulus and the loss (viscous) modulus.
For a generic stress wave $\sigma_{0}e^{i \omega t}$, the strain wave lagging behind  would be $\gamma_{0}e^{(i \omega t +\frac{\pi}{2})}$, hence the factor $i \equiv e^{i\frac{\pi}{2}}$ in front of $G''$, leading to the complex shear modulus defined as
\begin{equation}
    G^{*}=G' + i G''.
\end{equation}
In the linear response regime, with intrinsic material properties that do not vary with time and with causality being obeyed, $G'$ and $G''$ are related to each other by the Kramers-Kronig relations. The above relation applies equally to non-relativistic as well as to relativistic systems, as discussed below.

Furthermore, one obtains the following splitting of the shear stress into elastic and viscous components
\begin{equation}
    \sigma(t)=G'\epsilon_{xy}(t) + \frac{G''}{\omega}\dot{\epsilon}_{xy}(t) \label{fundamental}
\end{equation}
where $\epsilon_{xy}$ denotes the macroscopic shear strain.
For a relativistic off-diagonal shear deformation $\mu\nu=ik$ with $i \neq k$ the second term in the above equation represents the dissipative stress \cite{landau1959fluid}
\begin{equation}
\sigma' \equiv \sigma_{ik}'=c \eta \left(\frac{\partial u_{i}}{\partial x^{k}}+\frac{\partial u_{k}}{\partial x^{i}}-u_{k}u^{l}\frac{\partial u_{i}}{\partial x^{l}}-u_{i}u^{l}\frac{\partial u_{k}}{\partial x^{l}}\right)\label{relativistic_stress}
\end{equation}
and the off-diagonal relativistic strain rate:
\begin{equation}
\dot{\epsilon}_{xy} \equiv c\left(\frac{\partial u_{i}}{\partial x^{k}}+\frac{\partial u_{k}}{\partial x^{i}}-u_{k}u^{l}\frac{\partial u_{i}}{\partial x^{l}}-u_{i}u^{l}\frac{\partial u_{k}}{\partial x^{l}}\right)\label{relativistic_strain_rate}.
\end{equation}

In Eq. \eqref{fundamental}, the first term refers to the elastic part of the response,
which takes place on an infinitely long time-scale where relativistic effects are
absent, by construction. The second term in Eq. \eqref{fundamental} is the viscous stress 
described by Eqs. \eqref{relativistic_stress}-\eqref{relativistic_strain_rate} at the level of special relativity
(see \cite{landau1959fluid}). In particular, upon neglecting the elastic part (as appropriate for a nearly perfect fluid), the total
stress reduces to the viscous stress, $\sigma \approx \sigma'$. This, in turn, corresponds to an off-diagonal space-like component of the
stress-energy tensor, for example $\sigma = \sigma' = T^{12}$ where $T^{12}$ is the $\mu = 1, \nu = 2$
component of the stress-energy tensor $T^{\mu\nu}$ with $ \mu, \nu= 0, 1, 2, 3$. This component describes shear deformation in the spatial 12 plane of the Minkowski
space.

The above equations yield the following identification \cite{Landau_elasticity,landau1959fluid,lakes_2009}:
\begin{equation}
    \eta = \frac{G''}{\omega} \label{viscosity}
\end{equation}
between the viscosity $\eta$ of the system and the loss modulus $G''$.

We now follow the Langevin transport model for heavy quarks \cite{Teaney_2005,Teaney_2006}, and we generalize it by making it relativistic.
In the Langevin transport framework, the non-relativistic Langevin equation describes the motion of heavy quarks, including charm. One uses the fact that the heavy quark mass is much larger than the temperature scale, and that most of the heavy quarks in the QGP will have small momentum in the frame of the bath. The formalism is best suited for the bottom quark, mass $\sim 5$ GeV, and is also used for charm, mass $\sim 1.5$ GeV. Typical estimate of the temperature of QGP gives $\leq 0.3$ GeV.
A systematic expansion in $1/M$, the heavy quark mass, of the heavy quark motion has been worked out in Ref. \cite{Bouttefeux2020}. The size of the $1/M$ correction, following this formalism, was estimated in \cite{Banerjee2022} and was found to be moderate also for charm. Note, however, that this analysis is valid for heavy quarks moving slowly with respect to the heatbath frame (heavy quark momentum $\ll$ mass). The QGP itself is not non-relativistic, its main constituents are gluons and light quarks with mass $\ll$ temperature. So, in principle, the viscosity of QGP has to be evaluated in a relativistic framework, and the hydrodynamic formalism used for the medium flow has to be relativistic.

In the following we will estimate the relativistic viscosity contribution of the heavy quarks using a method valid for dense strongly-correlated matter, which provides a direct link between the shear viscosity and the frictional damping.

Introducing the mass-scaled tagged-particle displacement $s=Q\sqrt{m}$, with $Q$ the canonical coordinates, the resulting non-relativistic generalized Langevin equation of motion for the displacement of the tagged particle becomes (cfr. \cite{Zwanzig} for the full derivation from a microscopically reversible Caldeira-Leggett particle-bath Hamiltonian):
\begin{equation}
\ddot{s}=-U'(s)-\int_{-\infty}^t \nu(t-t')\frac{ds}{dt'}dt' + F_{p}(t),
\label{2.4gle1}
\end{equation}
where $F_{p}(t)$ is the thermal stochastic noise with zero average, $U$ is a local interaction potential (e.g. with the nearest neighbours), and $\nu$ is the friction resulting from many long-range interactions with all other particles in the system, imposed by the dynamical bi-linear coupling in the Caldeira-Leggett Hamiltonian \cite{Caldeira,Zwanzig,Petrosyan_2022,Zadra}.

As rigorously demonstrated in recent work \cite{Petrosyan_2022,Zadra}, the above Langevin equation can be written in manifestly covariant form for a relativistic system in terms of the four-momentum:
\begin{equation}
		\frac{d}{dt}p^\mu=F_\mathrm{ext}^\mu+F_p^\mu-\frac{1}{m}\int_0^t K{^\mu}_{\nu}(t,s) p^\nu(t-s)ds
		\label{eq:covlang}
\end{equation}
where $K{^\mu}_{\nu}(t,s)=\mathrm{diag}(0,K'^1(t,s),K'^2(t,s),K'^3(t,s))$ is the rank-two friction tensor \cite{Zadra}, and we have re-introduced the particle mass $m$.
Clearly, the non-relativistic and the relativistic Langevin equations, Eq. \eqref{2.4gle1} and Eq. \eqref{eq:covlang} respectively, have, \emph{mutatis mutandis}, the same form. Hence, from now on we shall work using the coordinates measured in the lab frame and employ suitable transformation laws when necessary.
The above relativistic Langevin equation can also be re-written in terms of the Cartesian 3-vector displacements of particle $i$ (measured in the deformed lab frame) \cite{Zadra}:
\begin{equation}
		\frac{d}{dt}(m\gamma\dot{\mathbf{r}}_i)=-\mathbf{F}_\mathrm{ext}+\mathbf{F}_p++\int \gamma\,\dot{\mathbf{r}}_i(t-s) \nu(t,s)ds,
	\label{eq:lang}
\end{equation}
where $\gamma$ is the Lorentz factor and we take an isotropic frictional memory-function, $K^{\mu}_{\nu}(t)\equiv \nu(t)$.

The relativistic velocity transformation between the lab frame and the co-moving frame of the macroscopic deformation is given by:
\begin{equation}
    \dot{\mathbf{r}}_{i} = \frac{1}{ 1 - \frac{\mathbf{u}\cdot\dot{\mathring{\mathbf{r}}}_i }{c^2} }\left[\frac{\dot{\mathring{\mathbf{r}}}_i }{\gamma_\mathbf{u}} - \mathbf{u} + \frac{1}{c^2}\frac{\gamma_\mathbf{u}}{\gamma_\mathbf{u} + 1}\left(\dot{\mathring{\mathbf{r}}}_i \cdot\mathbf{u}\right)\mathbf{u}\right]\label{transformation}
\end{equation}
where the dot indicates a time derivative (with respect to time $t$ measured in the lab frame) while the circle indicates quantities measured in the undeformed rest (lab) frame. Furthermore,
\begin{equation}
    \gamma_\mathbf{u} = \frac{1}{\sqrt{1-\dfrac{\mathbf{u}\cdot\mathbf{u}}{c^2}}}
\end{equation}
where $\mathbf{u}=\dot{\mathbf{F}}\mathbf{\mathring{r}}_{i}$ represents the local velocity of the deformed frame.

Since we are interested in the zero-shear rate limit of the shear viscosity (static shear viscosity), which is the parameter that enters the Navier-Stokes equation and the subsequent reduction to Euler equations for the QGP, then both $\dot{\mathbf{F}}$ and $\mathbf{u}$ are small and certainly much smaller than the speed of light $c$. Hence, with $\mathbf{u} \ll c$, Eq. \eqref{transformation} reduces to the (Galilean) transformation:
\begin{equation}
    \dot{\mathbf{r}}_{i} =\dot{\mathring{\mathbf{r}}}_i - \mathbf{u}.
\end{equation}
We can thus rewrite Eq. \eqref{eq:lang} for a tagged particle in $d$-dimensions, which moves with an affine velocity prescribed by the strain-rate tensor 
$\dot{\mathbf{F}}$ as follows:
\begin{equation}
\frac{d}{dt}(m\gamma\dot{\mathbf{r}}_i)=\mathbf{f}_i-\int_{-\infty}^t \nu(t-t')\,\gamma \left(\dot{\mathring{\mathbf{r}}}_i - \mathbf{u}\right) dt'.
\end{equation}
We now can expand the time derivative on the l.h.s. to obtain:
\begin{equation}
\frac{d}{dt}\left[m_i\gamma(\dot{\mathbf{r}}_i)\dot{\mathbf{r}}_i\right]=\ddot{\mathbf{r}}_i\gamma(\dot{\mathbf{r}}_i)+\frac{\gamma^3(\dot{\mathbf{r}}_i)}{c^2}\langle \dot{\mathbf{r}}_i,\ddot{\mathbf{r}}_i \rangle_3.\label{Zadra}
\end{equation}
For heavy quarks, which are often treated non-relativistically, as e.g. in Refs. \cite{Teaney_2005,Teaney_2006,Das_2022}, it is totally acceptable to assume that $\frac{\gamma^3(\dot{\mathbf{r}}_i)}{c^2} \rightarrow 0$, and therefore approximate Eq. \eqref{Zadra} as:
\begin{equation}
\frac{d}{dt}\left[m_i\gamma\dot{\mathbf{r}}_i\right]\approx \gamma \ddot{\mathbf{r}}_i.
\end{equation}

We then arrive at:
\begin{equation}
\gamma \ddot{\mathbf{r}}_i=\mathbf{f}_i-\int_{-\infty}^t\nu(t-t')\,\gamma\left(\dot{\mathring{\mathbf{r}}}_i - \mathbf{u}\right) dt'
\end{equation}
where $\mathbf{f}_i^\mu=-\partial{U}/\partial{\mathbf{r}}_i^\mu$ generalises the $-U'(s)$ to the tagged particle.
Furthermore, we used the following consideration.

This notation is consistent with the use of the circle on the particle position variables to signify that they are measured with respect to the reference rest frame (undeformed). In terms of this original rest frame $\lbrace\mathbf{\mathring{r}}_i\rbrace$, the equation of motion can be written, for the particle position averaged over several oscillations, as
\begin{equation}
\mathbf{F}\,\gamma \mathbf{\ddot{\mathring{r}}}_i  =\mathbf{f}_i-
\int_{-\infty}^{t}\nu(t-t')\cdot \,\gamma\frac{d\mathring{\mathbf{r}}_i}{dt'}dt',
\label{2.4restframe}
\end{equation}
where $\langle F_{p} \rangle=0$ is used upon averaging over many particles.

We work in the linear regime of small strain $\parallel\mathbf{F}-\mathbf{1}\parallel \ll 1$ by making a perturbative expansion in the small displacement $\{\mathbf{s}_i(t)=\mathbf{\mathring{r}}_i(t)-\mathbf{\mathring{r}}_i\}$ around a known rest frame $\mathbf{\mathring{r}}_i$.
That is, we take $\mathbf{F}=\mathbf{1}+\delta\mathbf{F}+...$ where $\delta\mathbf{F}\approx\mathbf{F}-\mathbf{1}$ is the small parameter. Replacing this back into Eq. \eqref{2.4restframe} gives
\begin{align}
&(\mathbf{1}+\delta\mathbf{F}+...)\,\gamma\frac{d^2\mathbf{s}_i}{dt^2}=\delta\mathbf{f}_i+\\
&-(\mathbf{1}+\delta\mathbf{F}+...)\int_{-\infty}^{t}\nu(t-t')\cdot\,\gamma\frac{d\mathbf{s}_i}{dt'}dt'.
\label{2.4expansion}
\end{align}
For the term $\delta\mathbf{f}_i$, imposing mechanical equilibrium again, which is $\mathbf{f}_{i}=0$,  
implies:
\begin{equation}
    \delta \mathbf{f}_{i} = \frac{\partial \mathbf{f}_{i}}{\partial \mathbf{\mathring{r}}_j }\delta \mathbf{\mathring{r}}_j + \frac{\partial \mathbf{f}_{i}}{\partial \mathbf{\eta}}:\delta \mathbf{\eta}
    \label{delta}
\end{equation}
where the double dot product is used because $\eta$ is a rank-two tensor.
In the first term on the r.h.s. of the above expression Eq. \eqref{delta} we recognise
\begin{equation}
    \frac{\partial \mathbf{f}_{i}}{\partial \mathbf{\mathring{r}}_j }\delta \mathbf{\mathring{r}}_j =-\mathbf{H}_{ij}\mathbf{s}_{j}
\end{equation}
where $\mathbf{H}_{ij}$ represents the Hessian matrix, defined as
\begin{equation}
\mathbf{H}_{ij}=\frac{\partial U}{\partial \mathring{\mathbf{r}}_{i} \partial \mathring{\mathbf{r}}_{j}}\bigg\rvert_{\gamma \rightarrow 0} = \frac{\partial U}{\partial \mathbf{r}_{i} \partial \mathbf{r}_{j}}\bigg\rvert_{\mathbf{r}\rightarrow \mathbf{r}_{0}}\label{ring_Hessian}
\end{equation}
since $\mathring{r}(\gamma)\rvert_{\gamma \rightarrow 0}=\mathbf{r}_{0}$, where $U$ denotes the total energy of the system.

For the second term in Eq. \eqref{delta} we have the following identification:
\begin{equation}
    \mathbf{\Xi}_{i,\kappa\chi}=\frac{\partial \mathbf{f}_{i}}{\partial \mathbf{\eta}_{\kappa\chi}}
\end{equation}
and the limit $\mathbf{\eta}_{\kappa\chi} \rightarrow 0$ is implied. Here $\eta_{\kappa\chi}$ are the components of the Cauchy-Green strain tensor defined as $\mathbf{\eta} =\frac{1}{2}\left(\mathbf{F}^{T}\mathbf{F}-\mathbf{1} \right)$. This is a second-rank tensor, and should not be confused with the fluid viscosity (a scalar), also denoted as $\eta$.
From a physical point of view, $\mathbf{\Xi}_{i,\kappa\chi}$ represents the force acting on a heavy quark in the affine position.

With these identifications, we can write Eq. (\ref{2.4expansion}), to first order:
\begin{equation}
\gamma\frac{d^2\mathbf{s}_i}{dt^2}+\int_{-\infty}^{t}\nu(t-t')
\,\gamma\frac{d\mathbf{s}_i}{dt'}dt'+\mathbf{H}_{ij}\mathbf{s}_{j}=\mathbf{\Xi}_{i,\kappa\chi}\eta_{\kappa\chi}.
\label{2.4gle2}
\end{equation}

The above equation can be solved by Fourier transformation followed by a normal mode decomposition, as we shall see next. If we specialise on time-dependent uniaxial strain along the $x$ direction, $\eta_{xx}(t)$, then the vector $\mathbf{\Xi}_{i,xx}\eta_{xx}(t)$ represents the force acting on particle $i$ due to the motion of its nearest-neighbors which are moving towards their respective affine positions (see e.g.~\cite{Lemaitre} for a more detailed discussion). Hence, all terms in the above Eq. \eqref{2.4gle2} are vectors in $\mathbb{R}^{3}$ and the equation is in manifestly covariant form.

To make it convenient for further manipulation, we extend all matrices and vectors to $Nd \times Nd$ and $Nd$-dimensional, respectively, and we will select $d=3$.
After applying Fourier transformation to Eq. \eqref{2.4gle2}, we obtain
\begin{equation}
-\gamma\,\omega^2\,\tilde{\mathbf{s}}+i\gamma\tilde{\nu}(\omega)\omega\,\tilde{\mathbf{s}}
+\mathbf{H}\,\tilde{\mathbf{s}}
=\mathbf{\Xi}_{\kappa\chi}\tilde{\eta}_{\kappa\chi},
\end{equation}
where $\tilde{\nu}(\omega)$ is the Fourier transform of $\nu(t)$ etc (we use the tilde consistently throughout to denote Fourier-transformed quantities). In the above equation, all the terms are are now vectors in $\mathbb{R}^{3N}$ space (we work in the heavy $M\rightarrow \infty$ mass limit, where $N$, the number of heavy quarks, is conserved).
Next, we apply normal mode decomposition in $\mathbb{R}^{3N}$ using the $3N$-dimensional eigenvectors of the Hessian as the basis set for the decomposition. This is equivalent to diagonalize the Hessian matrix $\mathbf{H}$. 
Proceeding in the same way as in \cite{Cui_viscoelastic}, we have that the $m$-th mode of displacement can be written as:
\begin{equation}
-\gamma\,\omega^2\hat{\tilde{s}}_m(\omega)+i\gamma\tilde{\nu}(\omega)\omega\,\hat{\tilde{s}}_m(\omega)
+\omega_m^2\hat{\tilde{s}}_m(\omega)
=\hat{\Xi}_{m,\kappa\chi}(\omega)\tilde{\eta}_{\kappa\chi}.
\label{2.4kerneldecom}
\end{equation}
It was shown \cite{Lemaitre}, by means of MD simulations that $\hat{\Xi}_{m,\kappa\chi}=\mathbf{v}_m\cdot\mathbf{\Xi}_{\kappa\chi}$ is self-averaging (even in the glassy state), and one might therefore introduce the smooth correlator function on eigenfrequency shells
\begin{equation}
\Gamma_{\mu\nu\kappa\chi}(\omega)=\langle\hat{\Xi}_{m,\mu\nu}\hat{\Xi}_{m,\kappa\chi}\rangle_{\omega_m\in\{\omega,\omega+d\omega\}} \label{Gamma}
\end{equation}
on frequency shells.
Following the general procedure of ~\cite{Lemaitre} to find the oscillatory stress for a dynamic nonaffine deformation, the stress is obtained to first order in strain amplitude as a function of $\omega$ (note that the summation convention is active for repeated indices):
\begin{align}
\tilde{\sigma}_{\mu\nu}(\omega)&=C^A_{\mu\nu\kappa\chi}\tilde{\eta}_{\kappa\chi}(\omega)-\frac{1}{\mathring{V}}\sum_{m}\hat{\Xi}_{m,\mu\nu}\hat{\tilde{s}}_m(\omega) \notag\\
&=C^A_{\mu\nu\kappa\chi}\tilde{\eta}_{\kappa\chi}(\omega)-\frac{1}{\mathring{V}}\sum_{m}\frac{\hat{\Xi}_{m,\mu\nu}\hat{\Xi}_{m,\kappa\chi}}{\omega_{m}^2-\gamma\omega^2
+i\gamma\tilde{\nu}(\omega)\omega}\tilde{\eta}_{\kappa\chi}(\omega)\notag\\
&\equiv C_{\mu\nu\kappa\chi}(\omega)\tilde{\eta}_{\kappa\chi}(\omega).\label{generalized}
\end{align}

In the thermodynamic limit and assuming a continuous vibrational spectrum, we can replace the discrete sum over $3N$ degrees of freedom with an integral over vibrational frequencies up to an ultraviolet cutoff frequency $\omega_{D}$ (in the kinetic theory of matter, this is known as the Debye frequency and is an increasing function of density $\omega_D \sim (N/\mathring{V})^{1/3}$). In this case, we need to replace the discrete sum over the $3N$ degrees of freedom (eigenmodes) with an integral, $\sum_{m=1}^{3N}...\rightarrow \int_{0}^{\omega_D}g(\omega_p)...d\omega_p$, where $g(\omega_p)$ is the density of states (DOS) or energy spectrum of the system, which can be normalized to unity or to the total number of degrees of freedom (e.g. $3N$) \cite{Kostya_book}.

Choosing the latter normalization of the DOS, the complex elastic constants tensor can be obtained as:
\begin{equation}
C_{\mu\nu\kappa\chi}(\omega)=C^A_{\mu\nu\kappa\chi}-\frac{1}{\mathring{V}}\int_0^{\omega_{D}}\frac{g(\omega_p)\Gamma_{\mu\nu\kappa\chi}(\omega_p)}{\omega_p^2-\gamma\omega^2+i\gamma\tilde{\nu}(\omega)\omega}d\omega_p
\label{2.4modulus}
\end{equation}
where $N/\mathring{V}$ denotes the density of the system in the initial undeformed state \cite{Cui_viscoelastic}. 
In the above we used mass-rescaled variables throughout, so that the particle mass is not present explicitly in the final expression.
If one, instead, uses non-rescaled variables and specializing to off-diagonal shear deformations, $\mu\nu\kappa\chi=xyxy$, we obtain the following relativistic expression for the complex shear modulus $G^{*}$:
\begin{equation}
G^{*}(\omega)=G_A-\frac{1}{\mathring{V}}\int_0^{\omega_{D}}\frac{g(\omega_p)\Gamma_{xyxy}(\omega_p)}{m\omega_p^2-m\gamma\omega^2+i\gamma\tilde{\nu}(\omega)\omega}d\omega_p.
\label{2.4shear_modulus}
\end{equation}
The first term on the r.h.s. is the affine shear modulus\index{affine shear modulus} $G_A$, which is independent of $\omega$.
The low-frequency behaviour of $\Gamma(\omega_p)$ can be estimated analytically using the following result:
\begin{equation}
\langle\hat{\Xi}_{p,\mu\nu}\hat{\Xi}_{p,\kappa\chi}\rangle = d\kappa
R_0^2\, \lambda_{p}\sum_{\alpha}B_{\alpha,\mu\nu\kappa\chi},\label{correlator}
\end{equation}
derived originally in Ref. \cite{Scossa}, which gives $\langle \hat{\Xi}_{p,xy}^{2}\rangle \propto \lambda_{p}$, thus implying (from its definition above): 
$\Gamma(\omega_p) \propto \omega_{p}^{2}$.
This analytical estimate appears to work reasonably well in the low-eigenfrequency part of the $\Gamma(\omega_p)$ spectrum of amorphous solids \cite{Palyulin,Lemaitre,Milkus2}. 
Furthermore, $\sum_{\alpha}B_{\alpha,\mu\nu\kappa\chi}$ is a geometric coefficient that depends only on the geometry of macroscopic deformation and its values can be found tabulated in Ref. \cite{Scossa}.

By separating real and imaginary part of the above expression, we then get to the storage and loss moduli as \cite{Milkus2}:
\begin{equation}
    \begin{split}
    G'(\omega)&=G_A - \frac{1}{\mathring{V}}\int_{0}^{\omega_{D}}\frac{m\,g(\omega_p)\,\Gamma(\omega_p)\,(\omega_{p}^{2}-\gamma\omega^{2})}{m^{2}(\omega_{p}^{2}-\gamma\omega^{2})^{2}+\gamma^2\tilde{\nu}(\omega)^{2}\omega^{2}}d\omega_p\\
    G''(\omega)&=\frac{1}{\mathring{V}}\int_{0}^{\omega_{D}}\frac{g(\omega_p)\,\Gamma(\omega_p)\,\gamma\,\tilde{\nu}(\omega)\,\omega}{m^{2}(\omega_{p}^{2}-\gamma\omega^{2})^{2}+\gamma^2\tilde{\nu}(\omega)^{2}\omega^{2}}d\omega_p.\label{visco-moduli}
    \end{split}
\end{equation}
It is easy to check that the storage modulus\index{storage modulus} $G'(\omega)$ reduces to the affine Born modulus $G_A \equiv G_\infty$ in the limit of infinite oscillation frequency, $\omega \rightarrow 0$.
From the point of view of practical computation, the DOS $g(\omega_p)$ can be obtained numerically via direct diagonalization of the Hessian matrix $\mathbf{H}_{ij}$, since its eigenvalues are related to the eigenfrequencies via $\lambda_p=m\omega_{p}^{2}$. Similarly, the affine-force correlator $\Gamma(\omega_p)$ can also be computed from its definition by knowing the positions of all the particles, their interactions and forces (so that the affine force fields $\mathbf{\Xi}$ can be computed) as well as the eigenvectors of the Hessian $\mathbf{v}_p$.

We note that, in the zero-frequency limit for the quasi-static elastic shear modulus we have:
\begin{equation}
    G'(0)=G_A - \frac{1}{\mathring{V}}\int_{0}^{\omega_{D}}\frac{\,g(\omega_p)\,\Gamma(\omega_p)\,}{m\omega_{p}^{2}}d\omega_p\\
\end{equation}
which agrees with previous derivations, cfr. \cite{Lemaitre}.
For a fluid or liquid at thermal equilibrium, it has been rigorously demonstrated that the two terms in the above equation (i.e. the positive affine term and the negative non-affine term) are identical, with opposite sign, see \cite{wittmer,zaccone2023}, resulting in $G'(0)=0$, in agreement with the well known fact that fluids and liquids possess zero shear rigidity and flow under an external mechanical perturbation. This also applies to the QGP which is well described by hydrodynamic equations of fluids, such as the Euler equation.

We recall that the viscosity can be obtained from the loss viscoelastic modulus $G''$ using nonaffine response theory by means of Eq. \eqref{viscosity}. 
The nonaffine response theory developed from first principles above thus provides the following form for the loss modulus $G''$ (cfr. Eq. \eqref{visco-moduli}):
\begin{equation}
G''(\omega)=\frac{1}{\mathring{V}}\int_{0}^{\omega_{D}}\frac{g(\omega_p)\,\Gamma(\omega_p)\,\gamma\,\tilde{\nu}(\omega)\,\omega}{m^{2}(\omega_{p}^{2}-\gamma\omega^{2})^{2}+\gamma^2\tilde{\nu}(\omega)^{2}\omega^{2}}d\omega_p. \label{loss_repeated}
 \end{equation}
For shear deformation ($\kappa\chi=xy$), 
Eq. \eqref{generalized} becomes:
\begin{equation}
    \sigma_{xy}(\omega)=G^{*}(\omega)\epsilon_{xy}(\omega).
\end{equation}
Since $G^{*}=G' + iG''$ and there is an overall factor $\omega$ in the above expression for $G''$ in Eq. \eqref{loss_repeated}, the theory correctly recovers, for the dissipative part of the stress, $\sigma_{xy}'$, the relativistic viscous flow given by Eq. \eqref{relativistic_stress}.
This, indeed, has a zero-frequency shear viscosity given by $\eta = \lim_{\omega \rightarrow 0}G''/\omega$, leading to:
\begin{equation}
    \eta = \frac{1}{\mathring{V}}\,\gamma\,\tilde{\nu}(0)\int_{0}^{\omega_{D}}\frac{g(\omega_p)\,\Gamma(\omega_p)}{m^{2}\omega_{p}^{4}}d\omega_p. \label{viscosity_nonaffine}
\end{equation}
As a sanity check, this formula correctly recovers the formula for the viscosity of non-relativistic fluids \cite{zaccone2023} upon taking the non-relativistic limit, i.e. as $\gamma \rightarrow 1$.

By comparing with the Kubo relationship used for QCD matter: $\eta = \lim_{\omega \rightarrow 0}\rho(\omega)/\omega$, with $\rho(\omega)$ the spectral function or spectral density \cite{kapusta_gale_2023}, we obtain the following identification:
\begin{equation}
    \rho(\omega) \sim G''(\omega).
\end{equation} 

Based on the most recent literature studies \cite{Datta_2022,Das_2022}, the low-energy part of the spectral function for the deconfined QGP can be written as \cite{Brambilla}: $\rho(\omega)=\frac{\omega\kappa_d}{2 T}$,
where $\kappa_d$ is the momentum diffusion coefficient. 



Since in the low-frequency sector the main excitations are acoustic waves with a linear dispersion, we take a Debye-type approximation for the DOS spectrum, $g(\omega_{p}) = \frac{\omega_{p}^2\,V}{2\,\pi^2\, c_s^3}$, where $c_s$ is the speed of sound.
Upon recalling Eq. \eqref{correlator} and $\lambda_p = m \omega_{p}^2$, the correlator $\Gamma(\omega_p)$ is given by $\Gamma(\omega_p) = \frac{1}{5} m\kappa R_{0}^{2} \omega_{p}^{2} $ \cite{Scossa}. By further including the degeneracy factor of 12 for the 3 colors, 2 spin states and particle-antiparticle, we thus get:
\begin{equation}
    \eta =  \frac{6}{5\pi^2c_s^3}\gamma\kappa R_{0}^{2} \,\tilde{\nu}(0) \frac{\omega_{D}}{m},\label{simplified2}
\end{equation}
where the effective spring constant $\kappa$ has dimensions of force per unit length, and in the high-T QGP, dominated by collisional physics, can be expressed in terms of the relevant energy scale divided by the separation distance squared, $\kappa \sim \frac{k_B T}{R_{0}^2}$. The rational behind this is that the interaction between heavy quarks will generate excitations of field quanta (gluons, or, for that matter, phonons), which eventually thermalize too.
For the same reason, in the above relation, $\omega_D \sim T$ (vibrational temperature, $\hbar \omega = k_B T$), and $\gamma \sim T$ for an ultrarelativistic fluid (since $\gamma m c^2 \approx k_B T$). Furthermore, in the regime $T>2 T_c$ we can assume the speed of sound $c_s$ of the QGP to be approximately $T$-independent \cite{BEGUN_2011}. Therefore, putting all these things together, we obtain the following estimate for the temperature dependence of the viscosity:
\begin{equation}
    \eta \sim T^3
\end{equation}
for the QGP viscosity in the high energy regime $T>2T_c$, which agrees very well with the available data and earlier calculations \cite{Redlich,Shuryak_RMP,kapusta_gale_2023}.

In conclusion, we have presented a microscopic derivation of the viscosity coefficient of strongly correlated dense QCD matter and we applied it to the case of  heavy quarks in the QGP within a new relativistic Langevin transport framework. 
The derivation leads to a fundamental expression for the viscosity contribution of the heavy quarks which turns out to be proportional to the cube of temperature, in agreement with previous studies. This provides an analytical microscopic derivation of QGP viscosity treated as a dense relativistic liquid. Furthermore, it provides a new fundamental link between the viscosity, the Lorentz factor, and the speed of sound. In future work, this model can be extended to include further microscopic details e.g. along the lines of \cite{Toneev}.

\subsection{Acknowledgments} 
I am indebted with Dr. Saumen Datta (Tata Institute of Fundamental Research, Mumbai) and to Dr. Lorenzo Gavassino (Vanderbilt University) for input and many useful discussions.

\bibliographystyle{apsrev4-1}

\bibliography{refs}

\end{document}